%Paper: hep-ph/9307243
%From: jenkins@su2xu1.UCSD.EDU (Elizabeth Jenkins)
%Date: Sat, 10 Jul 93 20:30:31 -0700

\input epsf
\input harvmac
\ifx\epsfbox\notincluded\message{(NO epsf.tex, FIGURES WILL BE IGNORED)}
\def\insertfig#1{}% null macro
\else\message{(FIGURES WILL BE INCLUDED)}
\def\insertfig#1{
\midinsert\centerline{\epsfxsize=4truein
\epsfbox{#1}}\endinsert}
\fi
%%%%%%%%%%%%%%%%%%%%%%%%%%%%%%%%%%%%%%%%%%%%%%%%%%%%%%%%%%%%%%%%%%%%%%
%
%  UCSD macros to overwrite some of the definitions in harvmac.tex
%  (include after harvmac.tex)
%  last modified 4/92
%
%%%%%%%%%%%%%%%%%%%%%%%%%%%%%%%%%%%%%%%%%%%%%%%%%%%%%%%%%%%%%%%%%%%%%%%
%
% modify the output routine for the little format
%
\ifx\answ\bigans
\else
\output={
  \almostshipout{\leftline{\vbox{\pagebody\makefootline}}}\advancepageno
}
\fi
%
%
% address
%

%
% grant numbers
%

%
% preprint number
%
\def\UCSD#1#2{\noindent#1\hfill #2%
\bigskip\supereject\global\hsize=\hsbody%
\footline={\hss\tenrm\folio\hss}}% restores pagenumbers
%
% abstract
%
\def\abstract#1{\centerline{\bf Abstract}\nobreak\medskip\nobreak\par #1}
%
%
% titlefont
%
%
\edef\tfontsize{ scaled\magstep3}
 \tfontsize  \tfontsize
 \tfontsize \font\titlei=cmmi10 \tfontsize
\font\titleis=cmmi7 \tfontsize \font\titleiss=cmmi5 \tfontsize
\font\titlesy=cmsy10 \tfontsize \font\titlesys=cmsy7 \tfontsize
\font\titlesyss=cmsy5 \tfontsize  \tfontsize
\skewchar\titlei='177 \skewchar\titleis='177 \skewchar\titleiss='177
\skewchar\titlesy='60 \skewchar\titlesys='60 \skewchar\titlesyss='60
%
%\def\titlefont{\def\rm{\fam0\titlerm}% switch to title font
%\textfont0=\titlerm \scriptfont0=\titlerms \scriptscriptfont0=\titlermss
%\textfont1=\titlei \scriptfont1=\titleis \scriptscriptfont1=\titleiss
%\textfont2=\titlesy \scriptfont2=\titlesys \scriptscriptfont2=\titlesyss
%\textfont\itfam=\titleit \def\it{\fam\itfam\titleit}\rm}
%
%
% math symbols
%
%---------------------------------------------------------------------
%
\def\inv{^{\raise.15ex\hbox{${\scriptscriptstyle -}$}\kern-.05em 1}}
  %prime
\def\lbar{{\lower.35ex\hbox{$\mathchar'26$}\mkern-10mu\lambda}} %lambda bar

%
%
% various slashed symbols
%
%
 % slashes a character
\def\dsl{\,\raise.15ex\hbox{/}\mkern-13.5mu D} %this one can be subscripted
\def\delsl{\raise.15ex\hbox{/}\kern-.57em\partial}
\def\Ksl{\hbox{/\kern-.6000em\rm K}}
\def\Asl{\hbox{/\kern-.6500em \rm A}}
\def\Dsl{\hbox{/\kern-.6000em\rm D}} %roman D
\def\Qsl{\hbox{/\kern-.6000em\rm Q}}
\def\gradsl{\hbox{/\kern-.6500em$\nabla$}}
%
% space and backspace in l mode
%
\def\lspace{\ifx\answ\bigans{}\else\qquad\fi}
\def\lbspace{\ifx\answ\bigans{}\else\hskip-.2in\fi} % $$\lbspace...$$
%
%     boxes an equation
%
\def\boxeqn#1{\vcenter{\vbox{\hrule\hbox{\vrule\kern3pt\vbox{\kern3pt
        \hbox{${\displaystyle #1}$}\kern3pt}\kern3pt\vrule}\hrule}}}
%
%     draw a little box (end of proof symbol)
%     e.g. \mbox{.1}{.1}
%
\def\mbox#1#2{\vcenter{\hrule \hbox{\vrule height#2in
\kern#1in \vrule} \hrule}}
%
%
%
%     curly letters
%
   %curly letters

  \def\CO{{\cal O}}

%
%
%
%     derivatives
%
%

%

\def\bar#1{\overline{#1}}

\def\bra#1{\left\langle #1\right|}
\def\ket#1{\left| #1\right\rangle}

\def\darr#1{\raise1.5ex\hbox{$\leftrightarrow$}\mkern-16.5mu #1}

%
 %pound sterling
%
 %puts a small half in a displayed eqn
\def\frac#1#2{{\textstyle{#1\over #2}}} %puts a small fraction
%in a displayed eqn
%
%
%     various math operators
%
%

%
%
%
%

%
%       relations
%
\def\ltap{\ \raise.3ex\hbox{$<$\kern-.75em\lower1ex\hbox{$\sim$}}\ }
\def\gtap{\ \raise.3ex\hbox{$>$\kern-.75em\lower1ex\hbox{$\sim$}}\ }
\def\gl{\ \raise.5ex\hbox{$>$}\kern-.8em\lower.5ex\hbox{$<$}\ }
\def\roughly#1{\raise.3ex\hbox{$#1$\kern-.75em\lower1ex\hbox{$\sim$}}}
%
%
%       This defines et al., i.e., e.g., cf., etc.

%

%
\def\np#1#2#3{{Nucl. Phys. } B{#1} (#2) #3}
\def\pl#1#2#3{{Phys. Lett. } B{#1} (#2) #3}

\def\physrev#1#2#3{{Phys. Rev. } {#1} (#2) #3}

\relax
%%
%% Young Tableaux
%%
%\def\lform{\hbox{$\sqcup$}\llap{\hbox{$\sqcap$}}}
\def\ssqr#1#2{{\vbox{\hrule height.#2pt
      \hbox{\vrule width.#2pt height#1pt \kern#1pt\vrule width.#2pt}
      \hrule height.#2pt}\kern-.#2pt}}
\def\sqr{\mathchoice\ssqr34\ssqr34\ssqr{2.1}3\ssqr{1.5}3}
%Sample tableaux:

\def\YtabB{\vbox{\hbox{$\sqr\sqr\sqr\thinspace$}\nointerlineskip
        \kern-.3pt\hbox{$\sqr$}}}
\def\YtabC{\vbox{\hbox{$\sqr\sqr\thinspace$}\nointerlineskip
        \kern-.3pt\hbox{$\sqr\sqr$}}}
\def\YtabD{\lower .5ex\hbox{\vbox{\hbox{$\sqr\sqr\thinspace$}
    \nointerlineskip\kern-.3pt\hbox{$\sqr$}
    \nointerlineskip\kern-.3pt\hbox{$\sqr$}}}}
\def\YtabE{\lower 1ex\hbox{\vbox{\hbox{$\sqr\thinspace$}
    \nointerlineskip\kern-.3pt\hbox{$\sqr$}
    \nointerlineskip\kern-.3pt\hbox{$\sqr$}
    \nointerlineskip\kern-.3pt\hbox{$\sqr$}}}}

\def\YtabG{\vbox{\hbox{$\sqr\sqr\thinspace$}\nointerlineskip
        \kern-.3pt\hbox{$\sqr$}}}
\def\YtabH{\lower .3ex\hbox{\vbox{\hbox{$\sqr\sqr\thinspace$}
    \nointerlineskip\kern-.3pt\hbox{$\sqr$}}}}
\def\YtabI{\lower 1ex\hbox{\vbox{\hbox{$\sqr\thinspace$}
    \nointerlineskip\kern-.3pt\hbox{$\sqr$}
    \nointerlineskip\kern-.3pt\hbox{$\sqr$}}}}
\def\YtabJ{\vbox{\hbox{$\sqr\thinspace$}\nointerlineskip
        \kern-.3pt\hbox{$\sqr$}}}

  \def\tinsert#1#2{\ooalign{$\hfil#1\mkern-.8mu
    \raise.105ex\hbox{$\times$}\hfil$\crcr$#1#2$}}

\noblackbox

\def\asl{\hbox{/\kern-.6500em A}}

\def\ssqr#1#2{{\vbox{\hrule height.4pt
      \hbox{\vrule width.4pt height#1pt \kern#1pt\vrule width.4pt}
      \kern-.5pt \hrule height.4pt}\kern-.4pt}}
\def\sqr{\ssqr{10}{10}}
\def\nbox{\hbox{$\sqr\sqr\sqr\sqr\cdot\cdot\cdot
\cdot\cdot\sqr\sqr\sqr\,$}}
\def\nboxA{\vbox{\hbox{$\sqr\sqr\sqr\sqr\cdot\cdot\cdot
\cdot\cdot\sqr\sqr\,$}\nointerlineskip
\kern-.3pt\hbox{$\sqr\,$}}}
\def\nboxB{\vbox{\hbox{$\sqr\sqr\sqr\sqr\cdot\cdot\cdot
\cdot\cdot\sqr\,$}\nointerlineskip
\kern-.3pt\hbox{$\sqr\sqr\,$}}}
\def\nboxD{\vbox{\hbox{$\sqr\sqr\cdot\cdot\cdot
\cdot\cdot\sqr\sqr\sqr\sqr\sqr\sqr\,$}\nointerlineskip
\kern-.2pt\hbox{$\sqr\sqr\cdot\cdot\cdot
\cdot\cdot\sqr\,$}}}
\def\nboxE{\vbox{\hbox{$\sqr\sqr\sqr\cdot\cdot\cdot
\cdot\cdot\sqr\sqr\sqr\sqr\,$}\nointerlineskip
\kern-.2pt\hbox{$\sqr\sqr\sqr\cdot\cdot\cdot
\cdot\cdot\sqr\,$}}}
\def\nboxF{\vbox{\hbox{$\sqr\sqr\sqr\sqr\cdot\cdot\cdot
\cdot\cdot\sqr\sqr\,$}\nointerlineskip
\kern-.2pt\hbox{$\sqr\sqr\sqr\sqr\cdot\cdot\cdot
\cdot\cdot\sqr\,$}}}
\def\nboxG{\vbox{\hbox{$\sqr\sqr\sqr\sqr\cdot\cdot\cdot
\cdot\cdot\sqr\,$}\nointerlineskip
\kern-.2pt\hbox{$\sqr\sqr\sqr\sqr\cdot\cdot\cdot
\cdot\cdot\sqr\,$}}}
\def\nboxH{\vbox{\hbox{$\sqr\sqr\cdot\cdot\cdot
\cdot\cdot\sqr\sqr\sqr\cdot\cdot\cdot\cdot\cdot\sqr\,$}\nointerlineskip
\kern-.2pt\hbox{$\sqr\sqr\cdot\cdot\cdot
\cdot\cdot\sqr\,$}}}
\def\clebsch#1#2#3#4#5#6{\left(\left.
\matrix{#1&#2\cr#4&#5\cr}\right|\matrix{#3\cr#6}\right)}
\def\threej#1#2#3#4#5#6{\left\{
\matrix{#1&#2&#3\cr#4&#5&#6\cr}\right\} }
\vskip 1.in
\centerline{{\titlefont{Light Quark Spin-Flavor
Symmetry}}}
\medskip
\centerline{{\titlefont{for Baryons Containing a Heavy
Quark}}}
\medskip
\centerline{{\titlefont{in Large N QCD}}}
\vskip .3in
\centerline{Elizabeth Jenkins}
\vskip .2in
\centerline{\sl Department of Physics}
\centerline{\sl University of California, San Diego}
\centerline{\sl 9500 Gilman Drive}
\centerline{\sl La Jolla, CA 92093}
\vfill
\abstract{The couplings and interactions
of baryons containing a heavy quark
are related by light quark spin-flavor symmetry
in the large $N$ limit.  The single pion coupling
constant which determines all heavy quark baryon-pion couplings
is equal to the
pion coupling constant for light quark baryons.  Light quark
symmetry relations amongst the baryon couplings are
violated at order $1/N^2$.  Heavy quark spin-flavor
symmetry is used in conjunction with large $N$ light
quark spin-flavor symmetry to determine the couplings
of the degenerate doublets of heavy quark baryons.
}
\vfill
\UCSD{\vbox{
\hbox{UCSD/PTH 93-17}\vskip-0.1truecm
\hbox{hep-ph/9307243}}}{June 1993}
%\draftmode
\eject

The large $N$ limit of QCD \ref\thooft{G. 't Hooft, \np {72}{1974}{461} }
originally led to a qualitative understanding
of baryon-pion interactions \ref\witten{E. Witten, \np
{160}{1979}{57}}.   Recent developments \ref\dm{R. Dashen and A.V.
Manohar, UCSD/PTH 93-16}
imply rigorous quantitative results.
In large $N$, it is possible
to show that baryon-pion couplings obey light quark spin-flavor
symmetry relations.  Furthermore, violation of these symmetry relations occurs
only at $\CO(1/N^2)$, since the $1/N$ correction has the
same group theoretic structure as the leading term \ref\rdam{R.
Dashen and A.V. Manohar, UCSD/PTH 93-18}.  Large $N$ relations for
other physical quantities can also be  obtained \dm\ref\jsq{E.
Jenkins, UCSD/PTH 93-19}. These results are identical to the
predictions of the Skyrme model in large $N$ \ref\adw{G.S. Adkins,
C.R. Nappi and E. Witten, \np {228}{1983}{552} }.  Since group
theoretic predictions of the Skyrme model are identical to the
predictions of the quark model in large $N$ \ref\csm{A.V. Manohar, \np
{248}{1984}{19} }\ref\gersak{J.-L. Gervais and B. Sakita, \physrev
{D30}{1984}{1795}}, these relations are also the same as those of the
non-relativistic quark model.  It is worth emphasizing, however, that
these results are model-independent and are rigorous consequences of
QCD in the large $N$ limit.

Another area of recent interest has been the study of baryons and
mesons which contain a single heavy quark.  In the heavy quark limit
$m_Q \rightarrow \infty$, the heavy quark effective lagrangian
possesses a heavy quark spin-flavor symmetry for a single heavy quark
with a given velocity \ref\iswis{N. Isgur and M.B. Wise, \pl {232}
{1989}{113}\semi \pl {237}{1990}{527}  }.  This heavy quark symmetry
can be used to derive relations between decay amplitudes and other
physical quantities for baryons or mesons containing a single heavy
quark.  For the study of baryons containing a single heavy quark,
it is possible to make use of both heavy quark symmetry and large $N$.
Recent work has studied the properties of baryons containing a single
heavy quark in the Skyrme model \ref\jmw{E. Jenkins, A.V. Manohar and
M.B. Wise, \np {396}{1993}{27} }. The heavy quark baryons arise in the
Skyrme picture as heavy meson-soliton bound states, where the solitons
are the ordinary baryons containing no heavy quark of the Skyrme model.
Quantum numbers and other properties of these heavy meson-soliton
bound states are in good agreement with experiment \ref\glm{Z.
Guralnik, M. Luke and A.V. Manohar, \np {390}{1993}{474}
}\ref\jmwtwo{E. Jenkins, A.V. Manohar and M.B. Wise, \np
{396}{1993}{38} }\ref\jm{E. Jenkins and A.V. Manohar, \pl
{294}{1992}{273} }.

The main emphasis of
this work is the study of pion-heavy quark baryon interactions.
in large $N$ QCD.
It is shown that a single pion coupling constant
parametrizes all heavy quark baryon-pion interactions.  More
importantly,     this single pion coupling constant is identical
to the single pion coupling constant describing baryon-pion interactions
for baryons containing no heavy quarks.  This argument can be generalized to
baryons containing any finite number of heavy quarks.  Thus, all pion-baryon
interactions are determined by a single pion coupling constant in large $N$
QCD.
As is the case for ordinary baryon-pion interactions, the large $N$
predictions for the heavy baryon-pion couplings respect light
quark spin-flavor symmetry.  Again, these symmetry relations are valid
to $\CO(1/N^2)$.

The large $N$ derivation of
baryon-pion couplings depends only on the isospin and angular momentum
assignments of the light degrees of freedom of the heavy
quark baryons, since the spin and flavor of the heavy quark are
conserved by low-energy QCD interactions.  The decoupling of the heavy
quark in the heavy quark $m_Q \rightarrow \infty$ limit makes the
analysis of pion couplings for heavy quark baryons identical to the
analysis for ordinary  baryons.
Heavy quark spin-flavor symmetry is used in conjunction with the large $N$
light quark spin-flavor symmetry to determine the couplings
of the degenerate doublets of heavy quark baryons.

Although this work concentrates on baryon-pion couplings, the large
$N$ arguments presented here have implications for other physical
parameters of the chiral lagrangian.   The general result of this work
is that physical parameters for heavy quark baryons are described by
an approximate light quark spin-flavor symmetry which becomes exact in
the large $N$ limit and an approximate heavy quark spin-flavor
symmetry which becomes exact in the heavy quark limit.  Together,
these two approximate symmetries greatly constrain chiral lagrangian
parameters.  In most cases, only a unique coupling remains for each
type of process.

The derivation of the above results begins with a discussion of the
spectrum of baryon states in the large $N$ limit.  For the case of
baryons containing no heavy quarks, the baryon spectrum for an odd
number of colors and $N_f = 2$ light flavors consists of  a degenerate
tower of isospin and angular momentum multiplets $(I,J)$ equal to
$(1/2, 1/2)$, $(3/2, 3/2)$, ...., $(N/2, N/2)$.  It is  possible to
show that this tower constitutes the minimal set of states which can
be present in large $N$ \dm.\footnote\dag{In principle,  there may be
additional towers of states degenerate with the minimal set. We will
simply assume the minimal choice.}   For the case of baryons
containing a single heavy quark, the minimal set of states consists of
a degenerate tower of $(I,J)$ states equal to $(0,0)$, $(1,1)$,
$(2,2)$, ...., $( (N-1)/2, (N-1)/2 )$, where $J$ represents the
angular momentum of the light degrees of freedom of the heavy quark
baryon.  (Note that each $(I,J)$ state in this tower corresponds to a
degenerate doublet of heavy baryon multiplets with isospin $I$ and
total spin equal to  $I\pm \frac 1 2$, since the angular momentum of
the light degrees of freedom $J=I$ and the spin of the heavy quark
$S_Q = \frac 1 2$.) Each of these two towers of states has the correct
quantum numbers to be generated from a single $SU(2 N_f)$ multiplet.
For the first tower, the $SU(4)$ representation is the totally
symmetric tensor product of $N$ fundamental representations of
$SU(4)$.  The Young tableau for this representation is a row of $N$
boxes. Under the breaking of the spin-flavor group to its isospin and
spin subgroups, $SU(4) \rightarrow SU(2) \times SU(2)$, this
representation produces the states of the first tower.   The tower of
states for the light degrees of freedom of heavy quark baryons  has
the correct quantum numbers for it to arise from the $SU(4)$
representation of $(N-1)$ completely symmetrized boxes.
Representative Young tableaux for the breaking of these
representations under the isospin and spin subgroups are given in
\fig\ytab{The towers of states $(I,J)$, $I=J$ equals half-integer or
$I=J$ equals integer arise from the totally symmetric $SU(4)$
representation $(a)$ of $N$ or $(N-1)$ boxes.  Under the breaking
$SU(4) \rightarrow SU(2) \times SU(2)$, the $SU(4)$ representation
breaks into isospin and spin $SU(2)$ representations of all Young
tableaux with two rows formed from $N$ or $(N-1)$ boxes.   The  two
highest dimensionsal representations are given by $(a)$ and $(b)$.  A
general representation has the form $(c)$.  The lowest
representation $(d)$ is the $(\frac 1 2, \frac 1 2)$ or $(0,0)$
state.}.

It is possible to prove that the
pion couplings of the multiplets in each degenerate tower of states
are determined by a single coupling constant. The group theoretic
structure of the couplings for each tower are the couplings which
result if the tower forms an $SU(4)$ representation.   We first review
the derivation of this result for ordinary baryon-pion interactions
\dm; then the argument is extended to heavy quark baryons.

General large $N$ considerations imply that pion-baryon scattering should be
$\CO(1)$ in the large $N$ expansion \witten.
Ref. \dm\ shows that this behavior only occurs if the there is an exact
cancellation amongst graphs at leading order in $N$, since for arbitrary
couplings
pion-baryon scattering will grow with $N$.  Thus, unitarity of
pion-baryon scattering in the large $N$ limit constrains pion-baryon
couplings.  As Ref.~\dm\ shows, this large $N$ constraint uniquely
determines all pion-baryon couplings in terms of a single coupling
constant.

Consider a
general baryon-pion vertex
\eqn\bpib{
\bar B_2 \,G^{ai} B_1 \ {{\partial^a \pi^i} \over f_\pi}\ ,
}
where $a=1,2,3$ labels the angular momentum channel of the $p$-wave
pion,  $i=1,2,3$ labels the isospin of the pion, and $G^{ai}$ is an
operator with unit spin and isospin.  Eq.~\bpib\ is written
in terms of static baryon fields of the
baryon-pion chiral lagrangian \ref\bary{E. Jenkins and A.V. Manohar,
\pl {255}{1990}{558} } in the baryon rest frame.  All pion couplings
amongst baryon multiplets in the degenerate tower which are allowed by
isospin and angular momentum conservation are present and all
couplings are unrelated.  Because the emission of a pion can only
change the isospin or spin of the initial baryon by one unit, only
diagonal couplings and nearest neighbor off-diagonal couplings between
multiplets in the tower are allowed.  The couplings allowed by isospin
and angular momentum conservation can be parametrized by the reduced
matrix elements of $G$,
\eqn\njjcouplings{\eqalign{
\bra{ I_2 I_{2z},J_2 J_{2z} } &G^{ai}  \ket{ I_1 I_{1z}, J_1 J_{1z} }
\cr &= N\,g(J_1, J_2) \, \sqrt{ { 2J_1+1 } \over { 2J_2+1 } }
\clebsch{I_1}{1}{I_2}{I_{1z}}{i}{I_{2z}}
\clebsch{J_1}{1}{J_2}{J_{1z}}{a}{J_{2z}} \ , \cr
}}
where $I_1 = J_1$
and $I_2 = J_2$ for the assumed tower of states and  $g(J_1, J_2)$ are
arbitrary couplings of $\CO(1)$.  An explicit factor of $N$ has been
factored out of the reduced matrix elements in Eq.~\njjcouplings\ to
keep all $N$ dependence manifest.  The explanation of this factor of
$N$ is given by the same power counting argument that implies that the
axial vector coupling constant of the nucleon $g_A \sim N$.   Large
$N$ considerations relate the couplings $g(J_1, J_2)$.   First
consider $\pi$ nucleon scattering, $\pi N \rightarrow \pi N$, in large
$N$.  The scattering occurs through direct and crossed diagrams, each
with $N$ and $\Delta$ intermediate states.  Thus, the scattering
depends on the couplings $g_{NN}$ and $g_{\Delta N}$.  Large $N$
counting of the diagrams is as follows.  Each diagram receives a
factor of $1/N$ from explicit factors of $f_\pi$ since each diagram
contains two pion vertices and $f_\pi \sim \sqrt{N}$.  The matrix
elements at the vertices give $\langle G^{ai} \rangle \langle
G^{bj}\rangle \sim \CO({N^2})$ so that the scattering amplitude is of
order $N$, unless the leading terms cancel.  This cancellation
requires $g_{\Delta N} = g_{NN} \equiv g $.  All other couplings are
determined recursively to equal $g$.  For example, $\pi \Delta
\rightarrow \pi N$ scattering involves the one additional coupling
$g_{\Delta \Delta}$. The constraint that the matrix element cancels at
leading order in $N$ implies $g_{\Delta \Delta} = g$.   Scattering of
$\pi \Delta$ to the $(5/2,5/2)$ baryon involves one additional
coupling, $g(\frac 3 2\, ,\frac 5 2)$.  Again, this coupling must
equal $g$ to produce a scattering amplitude which is $\CO(1)$.  Thus,
all the pion couplings $g(J_1, J_2)$ are determined recursively to be
equal to $g$,
\eqn\ncouplings{\eqalign{
\bra{ I_2 I_{2z}, J_2 J_{2z} }
&G^{ai}  \ket{ I_1 I_{1z}, J_1 J_{1z} } \cr &= N\, g \,
\sqrt{ { 2J_1+1 } \over { 2J_2+1 } }
\clebsch{I_1}{1}{I_2}{I_{1z}}{i}{I_{2z}}
\clebsch{J_1}{1}{J_2}{J_{1z}}{a}{J_{2z}} \ . \cr
}}
The coupling relations
\ncouplings\ are the relations obtained by grouping the degenerate
tower of baryon states into the totally symmetric $SU(4)$
representation.

A similar analysis can be performed for pion couplings to the
tower of states for the light degrees of freedom of the heavy quark
baryons. In the heavy quark limit, the angular momentum of light
degrees of freedom is separately conserved in low-energy QCD
interactions, so the analysis can proceed without considering the
heavy quark spin. For the tower of integral isospin and spin, a
diagonal pion coupling to the $(0,0)$ state is not allowed.  The
first coupling in the series connects the $(0,0)$ and the $(1,1)$
states.  Exact cancellation  of the $\CO(N)$ contribution to $\pi +
(0,0) \rightarrow \pi + (1,1)$ scattering relates this coupling to the
diagonal $(1,1)$ pion coupling.  Again, all  couplings can be
determined recursively in terms of this single coupling. The pion
coupling relations for this tower are
\eqn\lcouplings{\eqalign{ \bra{
I_2 I_{2z}, J_2 J_{2z} } &G^{ai}  \ket{ I_1 I_{1z}, J_1 J_{1z} } \cr
&= N \, g^\prime \, \sqrt{ { 2J_1+1 } \over { 2J_2+1 }
}  \clebsch {I_1}{1}{I_2}{I_{1z}}{i}{I_{2z}}
\clebsch {J_1}{1}{J_2}{J_{1z}}{a}{J_{2z}}, \cr
}}
where $g^\prime$ denotes the coupling constant, $G^{ai}$ is an operator with
unit spin and isospin, and the tower of states requires integral $I=J$.  As
before, these are the relations which
follow if one forms the totally symmetric $SU(4)$ representation of
$(N-1)$ boxes from the tower of states for the light degrees of
freedom of the heavy quark baryons.

Note that in considering
pion-heavy quark baryon scattering, we have ignored the heavy quark
entirely.
The couplings obtained above
are the couplings to the light degrees of freedom of the heavy quark
baryons.  In order to obtain the couplings of the heavy quark baryons,
one must construct tensor products of the light quark angular momentum
and the heavy quark spin.  For instance, the $(0,0)$ state corresponds
to the spin-$\frac 1 2$ $\Lambda_Q$.  All other states correspond to
degenerate doublets of heavy quark baryons.  For example, the $(1,1)$
state corresponds to the spin-$\frac 1 2$ $\Sigma_Q$ and the
spin-$\frac 3 2$ $\Sigma_Q^*$. The pion couplings to the heavy baryon
states can be easily derived using Eq.~\lcouplings\ for the matrix
elements of the light degrees of freedom and the spin decomposition of
the heavy baryon states \eqn\hbstates{
\ket{J J_z} = \sum_{J_{\ell\,z}, S_{Q\,z}} \ket{J_\ell, J_{\ell\,z}}
\ket{S_Q S_{Q\,z}}
\clebsch{J_\ell}{S_Q}{J}{J_{\ell\,z}}{S_{Q\,z}}{J_z},
}
where $J_\ell$ and $S_Q=\frac 1 2$ are the angular momentum of the light
degrees of freedom and the spin of the heavy quark, respectively.
The result of this computation is
\eqn\hbcouplings{\eqalign{
\bra{ I^\prime I_{z}^\prime, J^\prime J_{z}^\prime } &G^{ai}
\ket{ I I_{z}, J J_{z} } = N \, g^\prime \,(-1)^{1+I+S_Q+J^\prime}
\,(-1)^{I^\prime-I-1}\,(-1)^{J^\prime-J-1}\cr
&\sqrt{ { (2I+1) } \, { (2J+1) } }
\threej{1}{I}{I^\prime}{S_Q}{J^\prime}{J}
\ \clebsch {I}{1}{I^\prime}{I_{z}}{i}{I^\prime_{z}}
\clebsch {J}{1}{J^\prime}{J_{z}}{a}{J^\prime_{z}}, \cr
}}
where substitutions $J_\ell = I$ and $J_\ell^\prime = I^\prime$ have been
made and the quantity in curly braces is the $6j$ symbol.  For a single heavy
quark, $S_Q = \frac 1 2$.  The spectrum of heavy baryon states thus consists
of a degenerate tower of doublets with integer isospin $I$ and half-integral
total spin $J= I\pm 1/2$.

To prove that the pion coupling constants for baryon-pion
and heavy quark baryon-pion interactions are equal to each other,
one needs to consider the coupling of heavy quark baryons to
ordinary baryons and heavy quark mesons.  The lowest-lying multiplet
of mesons containing a single heavy quark is a degenerate doublet
which consists of a pseudoscalar meson $P_Q$ and a vector meson
$P_Q^*$.  For $Q= b$, these are the $\bar B$ and $\bar B^*$, whereas
for $Q= c$, they are the $D$ and $D^*$.
The light degrees of freedom of this multiplet carry
isospin and angular momentum quantum numbers $(\frac 1 2, \frac 1 2)$.
All the baryon $\rightarrow$ heavy baryon couplings to a heavy meson
can be parametrized in terms of the reduced matrix elements of an
operator $H$,
\eqn\hmhbbcouplings{\eqalign{ \bra{ I_2 I_{2z}, J_2
J_{2z} } &H^{ai}   \ket{ I_1 I_{1z}, J_1 J_{1z} } \cr
&= h(J_1, J_2) \,
\sqrt{ { 2J_1+1 } \over { 2J_2+1 } }
\clebsch {I_1}{\frac 1 2}{I_2}{I_{1z}}{i}{I_{2z}}
\clebsch {J_1}{\frac 1 2}{J_2}{J_{1z}}{a}{J_{2z}} \ , \cr
}}
where $H^{ai}$ is an operator with isospin and spin $\frac 1 2$
\footnote\dag{Note that an inverse factor of the $P^{(Q)}$ meson decay
constant $f_P$ is implicit in the definition of the matrix
elements Eq.~\hmhbbcouplings.  If this factor is not included, the
matrix elements $\langle H^{ai} \rangle$ must be redefined with a
factor of $\sqrt{N}$.  The heavy meson vertex is then given by $\langle
H^{ai} \rangle / f_P$, which is still $\CO(1)$.}.   The ordinary baryon
has half-integral $I_1=J_1$, whereas the light degrees of freedom of
the heavy quark baryon has integral $I_2=J_2$.  This formula can be
generalized by tensoring in the spin of the heavy quark, but the
analysis is more transparent if only the angular momentum of the light
degrees of freedom is used.  The matrix elements  $\langle H^{ai}
\rangle$ are $\CO(1)$.  A simple way to see that this is the correct
power of $N$ is to consider the scattering baryon + heavy meson
$\rightarrow$ baryon + heavy meson through an intermediate heavy quark
baryon.  This scattering is a low momentum transfer process; the large
heavy quark mass flows through the diagram since the intermediate
baryon is a heavy quark baryon.  There is no possible crossed diagram
for this process since the heavy quark mass cannot be properly routed
through the crossed diagram.  Thus, no cancellation is possible, and
the tree graph must be $\CO(1)$, which implies that each matrix
element is $\CO(1)$.  Alternatively, one can think of the heavy meson
+ baryon $\rightarrow$ heavy baryon coupling as replacing a quark in
the baryon with a heavy quark of the same color.  If the absorbed and
emitted heavy mesons attach to the same quark line in the baryon,
there are no factors of $N$ involved.  If one of the $\CO(N)$ other
lines is involved in the heavy meson emission, then there must be an
exchange of a gluon between the different quark lines by momentum
conservation.  Gluon exchange introduces a compensating factor of
$1/N$, so the scattering is $\CO(1)$.

The pion couplings of the two baryon towers can be related by studying
the amplitudes for
$\pi$ + heavy baryon $\rightarrow$ heavy meson + baryon.  This
process involves $g$, $g^\prime$ and $h(J_1,J_2)$.  The two tree
graphs which contribute are given explicitly in \fig\lpi{Feynman diagrams for
the scattering $\pi + \Lambda_Q \rightarrow P_Q^{(*)} + N$.  In leading order
in $N$, only the direct and crossed diagrams contribute.  The $(I,J)$ quantum
numbers of the light degrees of freedom of each particle are given.  Note that
the scattering is a low-momentum process which does not depend on the heavy
quark mass or spin;  the large heavy quark mass $m_Q$ and the heavy quark spin
$S_Q$ are conserved in each diagram.} for the special case $\pi + \Lambda_Q
\rightarrow P_Q^{(*)} + N$.
The graph in which the pion couples to the heavy meson is order
$1/N$ with respect to these two diagrams and can be ignored.  No large momentum
transfer is involved in this process since the heavy quark mass flows through
each diagram.  The typical momentum transfer is $m(\Lambda_Q)-m(P_Q^{(*)})-m(N)
\sim \CO(\Lambda_{QCD})$, which is $\CO(1)$ in both the $1/N$ and $1/m_Q$
expansions.  The direct channel graph involves an intermediate
heavy baryon state, whereas the crossed diagram involves
an intermediate baryon.  Since the pion-baryon couplings grow like $\sqrt{N}$,
these diagrams will result in a scattering amplitude of $\CO(\sqrt{N})$ unless
the
diagrams cancel exactly at leading order.  The condition for cancellation
is
\eqn\hbbscat{
\bra{I^\prime I^\prime_z J^\prime J^\prime_z} H^{bj} G^{ai}
\ket{I I_z J J_z}
-\bra{I^\prime I^\prime_z J^\prime J^\prime_z} G^{ai} H^{bj}
\ket{I I_z J J_z} = 0,
}
where summation over a complete set of intermediate states is implied.
For the first term, the sum is over all states allowed by isospin and
angular momentum conservation with integral $I=J$; for the second
term, the sum is over all allowed states with half-integral $I=J$.
Using Eqs.~\ncouplings, \hbcouplings, and~\hmhbbcouplings, this
constraint can be rewritten in terms of the couplings $g$, $g^\prime$
and $h(J_1 ,J_2)$.  For a given initial heavy baryon with spin of the
light degrees of freedom $J$, the final baryon can have spin $J \pm
\frac 1 2$ or $J \pm \frac 3 2$.  The recursion relation for each of
these channels are: \eqn\iplusthreehalf{ g^\prime(J,J+1) h(J+1,J+\frac
3 2) - h(J, J+\frac 1 2)  g(J+\frac 1 2,J+\frac 3 2)=0 }
for $J^\prime =J + \frac 3 2$,
\eqn\iplushalf{\eqalign{
&g^\prime(J,J) h(J,J+\frac 1 2) - h(J, J+\frac 1 2) g(J+\frac 1 2,J+\frac 1 2)
\cr &+ g^\prime(J,J+1) h(J+1,J+\frac 1 2) -h(J, J-\frac 1 2)
g(J-\frac 1 2,J+\frac 1 2) =0\cr
}}
for $J^\prime =J + \frac 1 2$,
\eqn\iminushalf{\eqalign{
&g^\prime(J,J) h(J,J-\frac 1 2) - h(J, J+\frac 1 2) g(J+\frac 1 2,J-\frac 1 2)
\cr &+ g^\prime(J,J-1) h(J-1,J-\frac 1 2) -h(J, J-\frac 1 2)
g(J-\frac 1 2,J-\frac 1 2) =0 \cr
}}
for $J^\prime =J - \frac 1 2$, and
\eqn\iminusthreehalf{
g^\prime(J,J-1) h(J-1,J-\frac 3 2) - h(J, J-\frac 1 2)
g(J-\frac 1 2,J-\frac 3 2)=0
}
for $J^\prime =J - \frac 3 2$, where $J$ is an integer.
The initial conditions are that $h(0, -\frac 1 2)=h(0, -\frac 3 2)=
h(1, -\frac 1 2) =0$.  In addition, all allowed pion couplings can be set equal
to $g$ or $g^\prime$.  Eq.~\iplusthreehalf\ implies
\eqn\recur{
h(J, J+ \frac 1 2) = \left( {g \over g^\prime} \right)^J h(0, \frac 1 2),
}
whereas Eq.~\iminusthreehalf\ implies
\eqn\recurtwo{
h(J, J- \frac 1 2) = \left( {g^\prime \over g} \right)^{J-1} h(1, \frac 1 2),
}
for all integer $J$.  Eq.~\iplushalf\ for $J=0$ (recall that
$g^\prime(0,0) = 0$) yields
\eqn\hrel{
h(1, \frac 1 2) = \left( {g \over g^\prime} \right) h(0, \frac 1 2).
}
Thus, if there is a consistent solution of the recursion relations,
all of the $h(J,J\pm \frac 1 2)$ couplings are given in terms of $h(0,
\frac 1 2)$. Finally, substituting Eqs.~\recur, \recurtwo, and \hrel\
into Eq.~\iplushalf, one obtains the constraint
\eqn\constraint{
( g^\prime - g)\left( {g \over g^\prime} \right)^J
+ g^\prime \left( {g^\prime \over g} \right)^{J-1}
-g \left( {g^\prime \over g} \right)^{J-2}=0,
}
which can only be satisfied for arbitrary $J$ if $g=g^\prime$.  Thus,
the coupling constants for the baryon-pion and heavy baryon-pion
interactions are equal and the heavy meson couplings of these baryons
are parametrized by a single coupling constant $h$,
\eqn\hcouplings{
\bra{ I_2 I_{2z}, J_2 J_{2z} } H^{ai}
\ket{ I_1 I_{1z}, J_1 J_{1z} }
= h \,
\sqrt{ { 2J_1+1 } \over { 2J_2+1 } }
\clebsch {I_1}{\frac 1 2}{I_2}{I_{1z}}{i}{I_{2z}}
\clebsch {J_1}{\frac 1 2}{J_2}{J_{1z}}{a}{J_{2z}} \ .
}
Note that there is no constraint
on the value of $h$.  The physical origin of this
coupling is quite different than that for $g$,
so no relation is to be expected.

The corrections to the spin-flavor relations of the pion couplings
for each tower of baryon states are order $1/N^2$ \rdam.  The relation
$g = g^\prime$ receives a correction at order $1/N$, however, since
the $1/N$ symmetry-preserving corrections to the two pion couplings
need not be equal.     This same power counting was obtained in the
soliton-heavy meson bound state model \glm.  The large $N$
approximation justifies these predictions.

The large $N$ arguments given in this work can be
generalized to baryons containing a finite number of heavy quarks.   In this
case, one treats the heavy quarks as a subsystem with total angular momentum
$J_Q$.  The heavy quark subsystem must be in a state which is totally
symmetric under the heavy quark spin-flavor symmetry.  Thus, the heavy
quark multiplets of baryons with a given number of heavy quarks form a
single representation of heavy quark spin-flavor symmetry, namely the
Young tableau with a row of $N_h$ boxes, where $N_h$ is the number of
heavy quarks in the baryon.  Low-energy QCD interactions do not couple
the heavy quark subsystem to the light degrees of freedom of baryon.
Thus, the analysis of pion couplings proceeds in terms of the flavor
and angular momentum quantum numbers of the light degrees of freedom
of the baryons as before.  The tower of states for the light degrees
of freedom of the baryons will be the half-integer (integer) $I=J$
towers for baryons with an even (odd) number of heavy quarks.
Unitarity of pion-baryon scattering amplitudes in large $N$ implies
that the pion couplings amongst baryons containing a given number of
heavy quarks are described by a single coupling constant.      Thus,
in the large $N$ limit, the tower of states for the light degrees of
freedom of the baryon forms the representation of light quark
spin-flavor symmetry given by the Young tableau with a row of
$(N-N_h)$ boxes. Pion-baryon scattering $\rightarrow$ heavy meson +
baryon identifies the pion coupling constants of baryons containing
$N_h$ heavy quarks and baryons containing $(N_h-1)$ heavy quarks.
(Note that this scattering must obey angular momentum conservation for
both the heavy degrees of freedom and the light degrees of freedom
involved in the process.)  Thus, large $N$ $SU(2N_f)$ light quark
symmetry in conjunction with $SU(2N_h)$ heavy quark symmetry implies
that all baryon-pion couplings are given in terms of a single coupling
constant up to corrections of order $1/N$ and $1/m_Q$.

\vfill\break\eject

\centerline{\bf Acknowledgements}
I thank R.~Dashen and A.V.~Manohar
for discussions.
This work was
supported in part by the Department of Energy
under grant number DOE-FG03-90ER40546.

\listrefs
\listfigs
\vfill
\eject

$$\nbox$$
\centerline{(a)}
\vskip.5in

$$\nboxA$$
\centerline{(b)}
\vskip.5in

$$\nboxH$$
\centerline{(c)}
\vskip.5in

$$\nboxF \quad\quad {\rm or}\quad\quad \nboxG$$
\centerline{(d)}
\vskip.5in

\centerline{Figure 1}
\vfill
\eject
\null\vskip 1.in
\insertfig{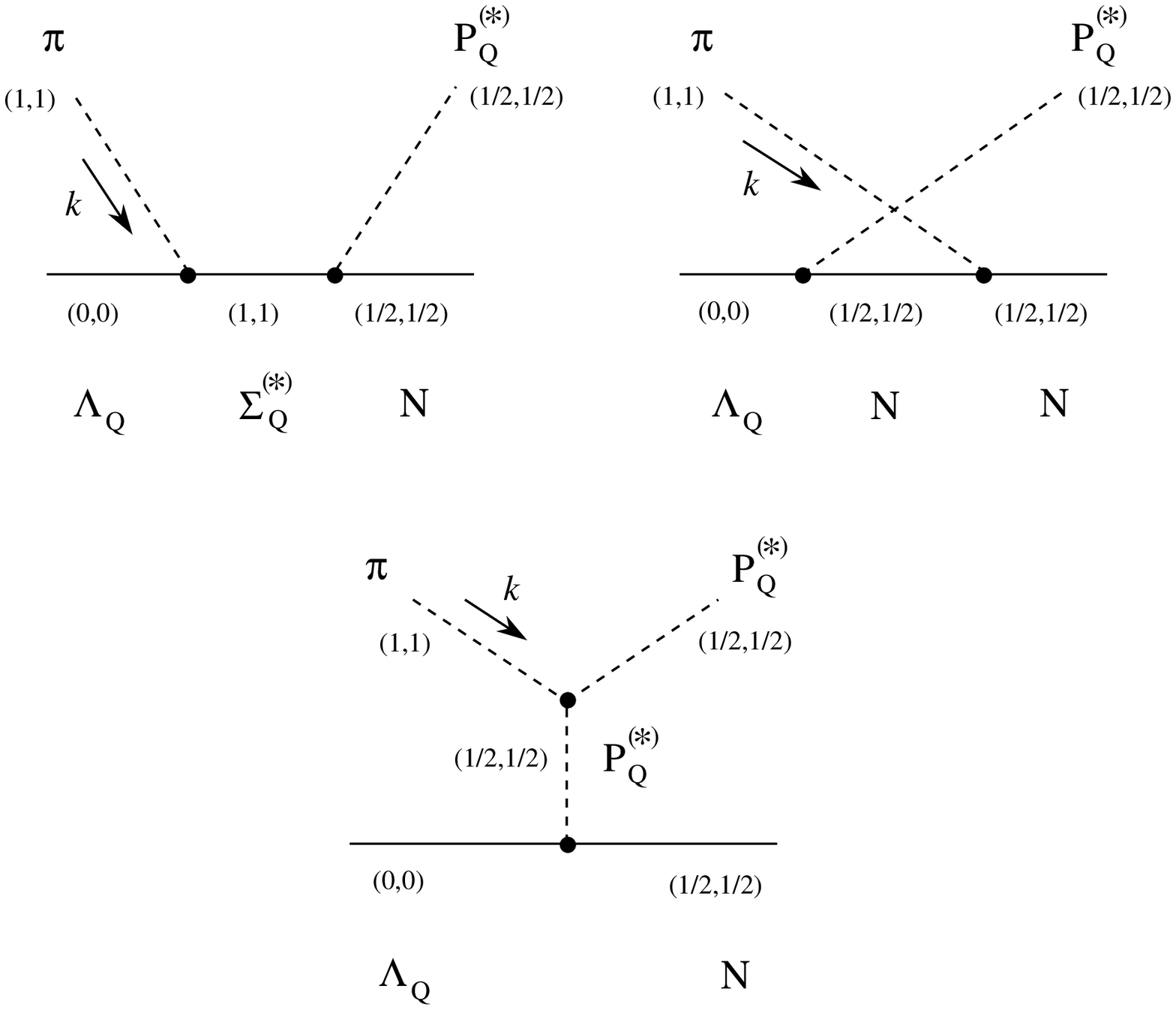}
\centerline{Figure 2}
\bye